\documentclass{article}
\usepackage{spconf,amsmath,epsfig}

\providecommand{\doi}[1]{doi: {\footnotesize \href{http://dx.doi.org/#1}{\path{#1}}}}

\usepackage[pdftex=true,breaklinks=true,hidelinks=true,colorlinks=true,citecolor=blue]{hyperref}

\usepackage[square,numbers]{natbib}
\title{Super-resolved rainfall prediction with physics-aware deep learning}

\twoauthors
  {S. Moran, B. Demir}
	{Technische Universit\"at Berlin\\
	Berlin, Germany}
  {F. Serva\sthanks{Now at the National Research Council of Italy}, 
  B. Le Saux}
	{European Space Agency $\Phi$-lab, ESRIN\\
	Frascati, Italy}

\begin{document}
%\ninept
%
\maketitle

\begin{figure*}[t]
\begin{center}
\includegraphics[width=1.55\columnwidth]{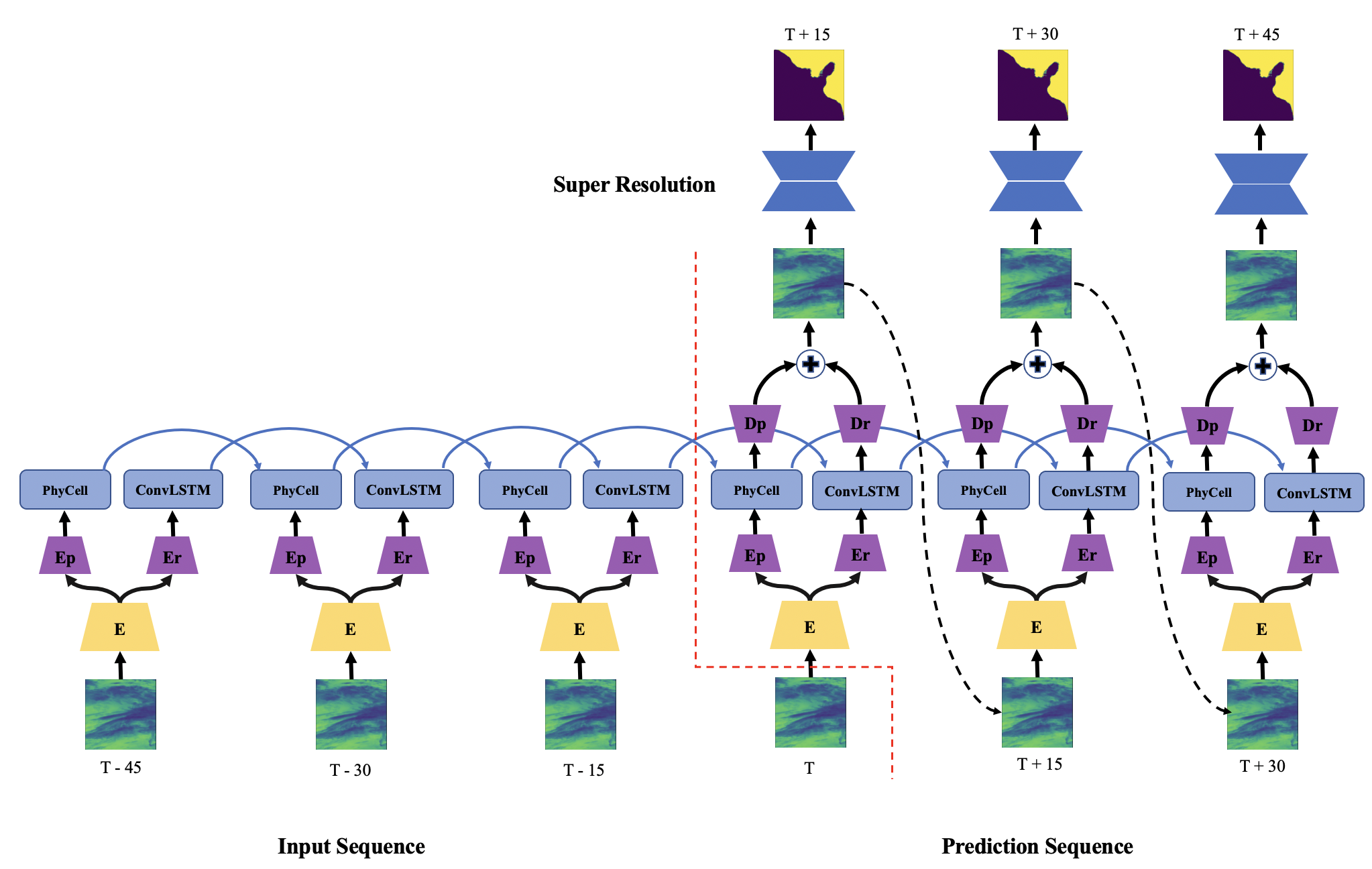}
\caption{Overview of the proposed SR-PhyDNet architecture for satellite image sequence prediction,
super-resolution and segmentation tasks. Satellite images are the 
input and a binary (rain in yellow, no rain in purple) mask is the output.}
\label{fig:sr-phydnet}
\end{center}
\end{figure*}

\begin{figure*}[t]
\begin{center}
\includegraphics[width=1.55\columnwidth]{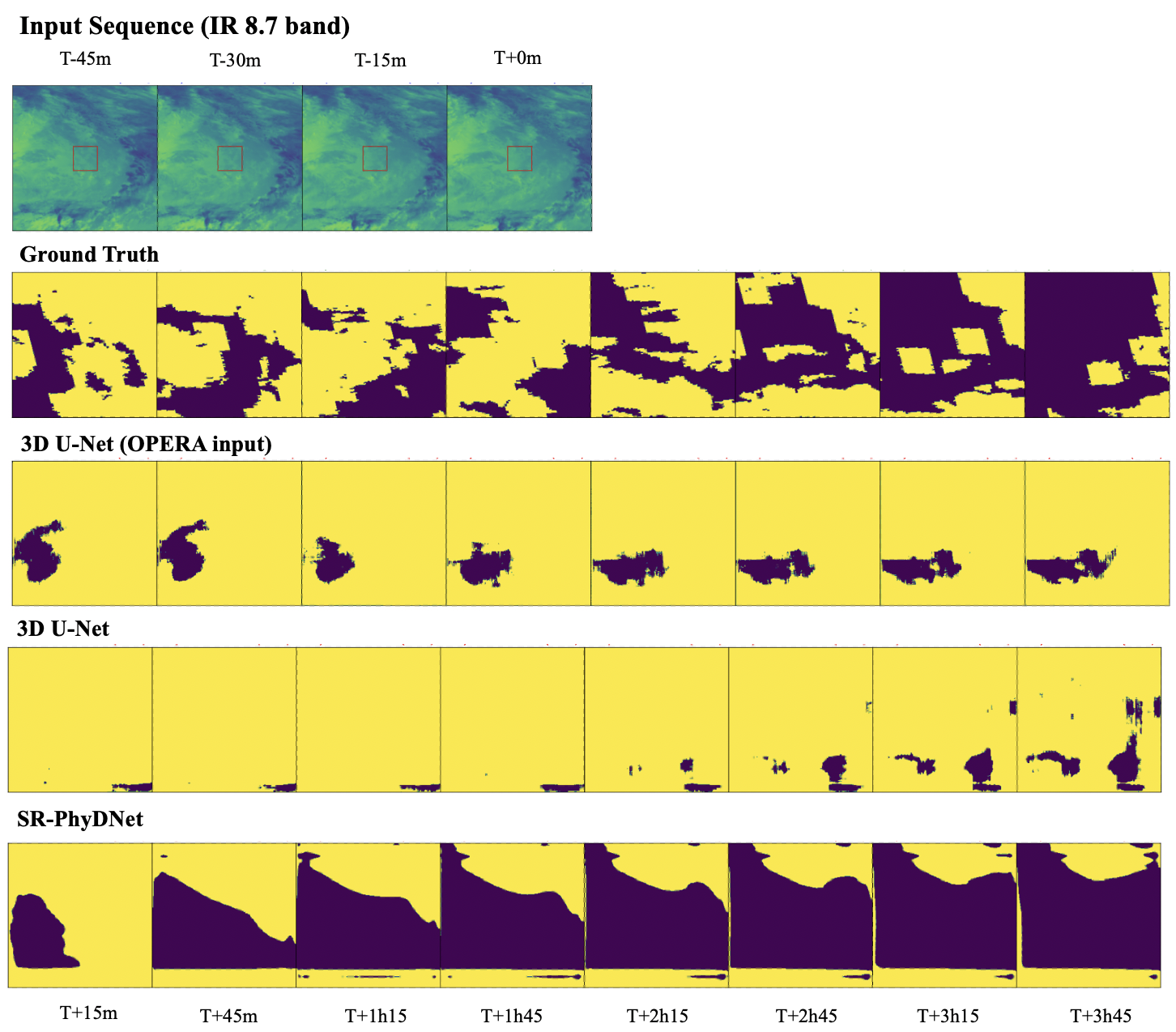}
\caption{Input satellite data (top, for one arbitrary infrared band) and comparison of different models with radar ground truth
for a heavy rain event dissipating over time (in rows 2-5, rain in yellow, no rain in purple).}
\label{fig:ex-heavy}
\end{center}
\end{figure*}

\begin{abstract} % 100-150 words
Rainfall prediction at the kilometre-scale up to a few hours in the future is key for 
planning and safety. But it is challenging given the complex influence of climate change on cloud processes and the limited skill of weather models at this scale.  
Following the set-up proposed by the \emph{weather4cast} challenge of NeurIPS, 
we build a two-step deep-learning solution for predicting rainfall occurrence
at ground radar high spatial resolution starting from coarser resolution weather 
satellite images.
Our approach is designed to predict future satellite images with a physics-aware
ConvLSTM network, which is then converted into precipitation maps through a U-Net.
We find that our two-step pipeline outperforms the baseline model
and we quantify the benefits of including physical information. 
We find that local-scale rainfall predictions with good accuracy starting 
from satellite radiances can be obtained for up to 4 hours in the future.
\end{abstract}
\section{Introduction}
\label{sec:intro}

Improvements in remote sensing technologies and the growth of computational capabilities
in the last decades have allowed us to predict the weather with higher skill compared to the past \cite{bauer_quiet_2015}. At the core of current weather forecast systems are complex weather models referred to as Numerical Weather Prediction (NWP) methods. Providing accurate forecasts of precipitation events is important given their impacts on society and our economies~\cite{Rambour-isprs2020-flood}.

NWP models are computationally expensive, therefore current research \cite{schultz_can_2021, bonavita_machine_2021, esaecmwfreport2021} is focusing on the potential for Deep Learning (DL) methods especially for 
short-term forecasting (nowcasting, up to a few hours in the future, \cite{sun_use_2014}).
State-of-the-art DL methods have good accuracy for low rain rates but provide blurry 
predictions for longer lead times and more impactful medium and heavier rainfall rates~\cite{ravuri_skillful_2021}.

Including physical constraints is a possible solution for more robust predictions while also taking advantage of DL to learn complex relationships from the data~\cite{mcgovern_making_2019, schultz_can_2021}. While weather satellites provide high-frequency imagery at a moderate spatial resolution, ground-based radars give precipitation estimates with high spatial resolution but have limited coverage due to their high costs. In \cite{sonderby_metnet_2020, espeholt_skillful_2021, ravuri_skillful_2021}
it has been shown that combining radar rainfall estimates with weather satellite data for training 
large DL networks can be competitive with NWP models.

In this work, we tackle the \emph{weather4cast} challenge prediction task by means of a physics-aware DL workflow to predict rainfall with high spatial resolution using only coarser resolution weather satellite imagery as input.

\section{Data and methods}

\subsection{Experiments setup}
 The problem formulation follows the \emph{weather4cast}\footnote{\url{https://www.iarai.ac.at/weather4cast/}, 2022 edition} competition (stage 1) presented at NeurIPS 2022: given a one-hour input sequence of Meteosat satellite 
 data~\cite{battrick_meteosat_1999} across the visible, infrared and water vapour bands, 
 binary precipitation maps (threshold 0.001 mm/hour) at the resolution 
 of OPERA ground-based radar~\cite{saltikoff_opera_2019} are to be predicted for the next hours every 15 mins.
 Since the spatial resolution of radar data is {\bf{6}} times finer than that of satellite images, the problem  requires a multi-modal sequence prediction task with super-resolution (SR).
 The input satellite sequences have the shape: channel, time, height, width (11, 4, 252, 252), while outputs are shaped as time, height, width 
 (16, 252, 252).
 To inform models about surrounding weather conditions, the satellite context covers 1008 km$^2$ on the ground, 
 while the target region is 168 km$^2$ large, i.e. six times smaller.
 
 We considered the three target regions of \emph{weather4cast} stage 1, but given the limited availability of resources we chose to train and test on a northern European region (\emph{b15}, see \cite{gruca_weather_2022}), 
 an area relatively balanced in terms of rain/no-rain distribution.
 Given the nowcasting focus, predictions are sought up to 4 hours in the 
 future.
 The evaluation metrics we use are the Recall, Precision, F1-score and Critical Success Index (CSI) or Intersection over Union (IoU). 

\subsection{Baseline 3D convolutional model}
A 3D U-Net architecture \citep{ronneberger_u-net_2015} is used as a baseline that can be trained 
end-to-end for this challenge. The model has a depth of 5 blocks. In each block, convolutional layers with 3D kernels are used to model spatio-temporal data relationships. Each block uses a Rectified Linear Unit (ReLU) activation function and dropout (0.4 probability) is added to prevent overfitting. The intermediate output size of the network is (32, 4, 252, 252), and a further convolutional layer and two further transpose convolutions are used to reshape outputs as tensors (16, 252, 252). 

\subsection{A PhyDNet/U-Net architecture}
Our physics-based DL solution for this task is the SR-PhyDNet architecture. As shown in Fig.~\ref{fig:sr-phydnet}, it consists of two sub-networks; 1) PhyDNet, a physics-based architecture described 
by \cite{guen_disentangling_2020}, used to predict the temporal evolution of satellite radiances, and 2) 
SR U-Net for the purpose of SR and image translation. The two sub-modules were trained independently and their combination makes the SR-PhyDNet a two-step workflow. PhyDNet is a two-branch recurrent architecture where one branch is a convolutional/recurrent ConvLSTM cell that models residual factors and the other is PhyCell which performs partial differential equation (PDE) constrained prediction with convolutional filters, providing physical consistency. The method is detailed in \cite{le_guen_deep_2020} and learns a latent space \textbf{H} where physics and residual factors are disentangled $h = h^p + h^r$ and follows the dynamics: $\partial_t h(t,x)= \partial_t h^p + \partial_t h^r$. 
The SR task is carried out by a U-Net sub-module, with a depth of 4, that converts PhyDNet satellite predictions into radar binary rain maps,
and performs a translation from the satellite context to a smaller target patch. 

Our approach was developed without considering the submissions to \emph{weather4cast}, but interestingly the winning team adopted a similar approach \cite{gruca_weather_2022}.
However, the winning solution uses a three-module architecture, where a sat2rad U-Net is used to convert the input sequence into a rainfall mask which is provided as additional input to the final predictive U-Net. Another difference is that in this approach the full context region is maintained throughout all predictions and only at the end is the image cropped and up-scaled to match the target resolution.
As a new contribution, in this work we quantify the impact of including physical information through ablation experiments.

\section{RESULTS AND ANALYSIS}
In this section, we report the overall classification performance of the methods described above.
Both the baseline 3D U-Net and the SR-PhyDNet models perform very well for non-rain events. The 3D U-Net shows some artefacts in its predictions but overall the number of FP is small. 
For rain events, as shown in Table~\ref{tab:comp_methods}, SR-PhyDNet outperforms the 3D U-Net by 3.2\% in terms of IoU. The drivers behind these metrics can be seen when the prediction sequences of each method are visualised. SR-PhyDNet achieves better scores and visually has fewer artefacts than the baseline as shown in Fig.~\ref{fig:ex-heavy}. For this precipitation event dissipating over 4 hours, the baseline results are generally more static than the physics-aware model. On the other hand, SR-PhyDNet tends to produce rounded-edged patterns likely due to overly strong diffusion at the edges. The predictions of the 3D U-Net contain straight lines and box-like artefacts in many instances. The use of OPERA input (a simpler task) provides more accurate predictions spatially for early lead times and reduced presence of artefacts, but cannot be extended to all zones covered by the satellite.

The metrics in Table \ref{tab:comp_methods} support this analysis. While the baseline method has a better test recall and can therefore correctly identify more positive instances, it has a lower precision than that of SR-PhyDNet and is, therefore, more prone to overestimate the extent of rain events. 
The inclusion of physical information improves predictions, which are up to 20-30\% better compared to when the PhyCell is bypassed (NoPhys).

As a control experiment, the baseline architecture with only OPERA data as input (3D U-Net-OPERA) is tested to understand better the effect of using satellite input for the task. This method obtains the best overall performances as expected with an IoU of 0.434 (Table \ref{tab:comp_methods}). However, temporal evolution similarly to the baseline remains too static (Fig.~\ref{fig:ex-heavy}). 

\begin{figure}[htbp]
\begin{center}
\includegraphics[width=.99\columnwidth]{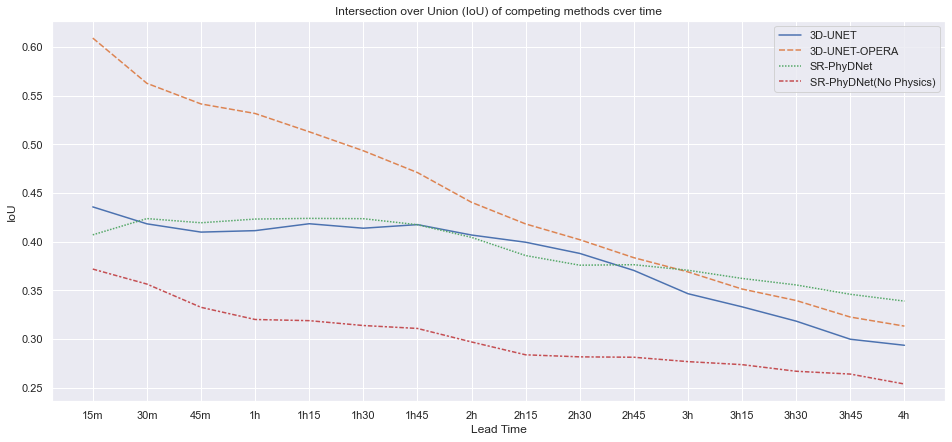}
\caption{ IoU metrics over a 4-hour forecast period. Higher values indicate better predictions.}
\label{fig:iou_ot}
\end{center}
\end{figure}

\begin{table}
\small
  \caption{Comparison of proposed methods on the classification task. Metrics 
  are averaged across test predictions. Best performing results among the first three are bolded.}
  \label{tab:comp_methods}
  \centering
  \begin{tabular}{lllll}
    \hline
    Method    & IoU\slash CSI & F1 & Recall & Prec \\
    \hline
    3D U-Net  & 0.378 & 0.549 & 0.808 &  0.416\\
    SR-PhyDNet & \textbf{0.391} & \textbf{0.562} & 0.771 & \textbf{0.442} \\
    SR-PhyDNet (NoPhys) & 0.300 & 0.461 & \textbf{0.832} & 0.319\\
    \hline
    3D U-Net (OPERA) & 0.434 & 0.606 & 0.890 & 0.459 \\
    \hline
  \end{tabular}
\end{table}
Fig.~\ref{fig:iou_ot} shows that using OPERA data as input makes the task easier for shorter lead times (3D U-Net-OPERA). 
Using satellite-only input however, SR-PhyDNet performs better than 3D U-Net baseline and we show that 
including physics helps on longer times, beyond 2-3 hours lead times.
These results are consistent with the findings of~\cite{pirht_weather_2022}, who however did not 
quantify the specific contribution from the PhyCell. 
Besides the quantitative metrics, predictions of SR-PhyDNet appear more realistic visually,
which is important for applications \cite{ravuri_skillful_2021}.

\section{Conclusions and future work}
In this work, we propose a physics-aware DL workflow for predicting precipitation events 
with the spatial resolution of ground-based weather radars starting from coarser resolution satellite imagery. Following the \emph{weather4cast} problem formulation, we study the advantage given by a combination of ConvLSTM networks with physics-aware (through PDE prediction) components to predict satellite-like images of cloud evolution over time when compared to a baseline 3D U-Net architecture for a region with a balanced rain/no-rain distribution. 

Our results show that while not a replacement for weather radar rain measurements, 
satellite data-driven deep learning models are suitable for predicting precipitation 
in areas lacking ground-based instruments \cite{hill_skilful_2020}.
This should be tested with additional experiments over different regions and climate conditions (as done in \cite{gruca_weather_2022}).
As a novel contribution, we demonstrate the importance of physical constraints through the superior 
performances and more realistic patterns of the SR-PhyDNet workflow, which
outperforms for longer lead times the baseline model even when precipitation is added in input.

% BiBTeX files 
% -------------------------------------------------------------------------

\bibliographystyle{plainnat}

\small

\bibliography{biblio}

\begin{thebibliography}{18}
\providecommand{\natexlab}[1]{#1}
\providecommand{\url}[1]{\texttt{#1}}
\expandafter\ifx\csname urlstyle\endcsname\relax
  \providecommand{\doi}[1]{doi: #1}\else
  \providecommand{\doi}{doi: \begingroup \urlstyle{rm}\Url}\fi

\bibitem[Battrick(1999)]{battrick_meteosat_1999}
B.~Battrick, editor.
\newblock \emph{Meteosat second generation: the satellite development}.
\newblock Number 153 in {ESA} {BR}. ESA Publications, Noordwijk, 1999.
\newblock ISBN 978-92-9092-634-4.

\bibitem[Bauer et~al.(2015)Bauer, Thorpe, and Brunet]{bauer_quiet_2015}
P.~Bauer, A.~Thorpe, and G.~Brunet.
\newblock The quiet revolution of numerical weather prediction.
\newblock \emph{Nature}, 525\penalty0 (7567):\penalty0 47--55, 2015.
\newblock \doi{10.1038/nature14956}.

\bibitem[Bonavita et~al.(2021)]{bonavita_machine_2021}
M.~Bonavita et~al.
\newblock Machine {Learning} for {Earth} {System} {Observation} and
  {Prediction}.
\newblock \emph{BAMS}, 102\penalty0 (4):\penalty0 E710--E716, 2021.
\newblock \doi{10.1175/BAMS-D-20-0307.1}.

\bibitem[Espeholt et~al.(2021)]{espeholt_skillful_2021}
L.~Espeholt et~al.
\newblock Skillful {Twelve} {Hour} {Precipitation} {Forecasts} using {Large}
  {Context} {Neural} {Networks}.
\newblock \emph{arXiv:2111.07470 [physics]}, 2021.

\bibitem[Gruca et~al.(2023)]{gruca_weather_2022}
A.~Gruca et~al.
\newblock Weather4cast at {NeurIPS} 2022: Super-resolution rain movie
  prediction under spatio-temporal shifts.
\newblock In \emph{PMLR Proc. {NeurIPS} Comp. Track}, volume 220, 2023.
\newblock URL \url{https://proceedings.mlr.press/v220/gruca22a.html}.

\bibitem[Hill et~al.(2020)]{hill_skilful_2020}
P.~G. Hill et~al.
\newblock How skilful are nowcasting satellite applications facility products
  for tropical {Africa?}
\newblock \emph{Meteorol Appl.}, 27:e1966, 2020.
\newblock \doi{10.1002/met.1966}.

\bibitem[{Le Guen} and Thome(2020{\natexlab{a}})]{guen_disentangling_2020}
V.~{Le Guen} and N.~Thome.
\newblock Disentangling {Physical} {Dynamics} from {Unknown} {Factors} for
  {Unsupervised} {Video} {Prediction}.
\newblock \emph{arXiv:2003.01460 [cs]}, 2020{\natexlab{a}}.

\bibitem[{Le Guen} and Thome(2020{\natexlab{b}})]{le_guen_deep_2020}
V.~{Le Guen} and N.~Thome.
\newblock A {Deep} {Physical} {Model} for {Solar} {Irradiance} {Forecasting}
  with {Fisheye} {Images}.
\newblock In \emph{2020 {IEEE}/{CVF} {Conf.} ({CVPRW})}, pages 2685--2688,
  2020{\natexlab{b}}.
\newblock \doi{10.1109/CVPRW50498.2020.00323}.

\bibitem[McGovern et~al.(2019)]{mcgovern_making_2019}
A.~McGovern et~al.
\newblock Making the {Black} {Box} {More} {Transparent}: {Understanding} the
  {Physical} {Implications} of {Machine} {Learning}.
\newblock \emph{Bulletin of the American Meteorological Society}, 100, 2019.
\newblock \doi{10.1175/BAMS-D-18-0195.1}.

\bibitem[Pihrt et~al.(2022)]{pirht_weather_2022}
J.~Pihrt et~al.
\newblock Weatherfusionnet: Predicting precipitation from satellite data.
\newblock \emph{arXiv:2211.16824 [cs]}, 2022.

\bibitem[Rambour et~al.(2020)]{Rambour-isprs2020-flood}
C~Rambour et~al.
\newblock Flood detection in time series of optical and sar images.
\newblock In \emph{ISPRS Archives}, volume XLIII, pages 1343--1346, 2020.
\newblock \doi{10.5194/isprs-archives-XLIII-B2-2020-1343-202}.

\bibitem[Ravuri et~al.(2021)]{ravuri_skillful_2021}
S.~Ravuri et~al.
\newblock Skillful {Precipitation} {Nowcasting} using {Deep} {Generative}
  {Models} of {Radar}.
\newblock \emph{Nature}, 597\penalty0 (7878):\penalty0 672--677, 2021.
\newblock \doi{10.1038/s41586-021-03854-z}.

\bibitem[Ronneberger et~al.(2015)Ronneberger, Fischer, and
  Brox]{ronneberger_u-net_2015}
O.~Ronneberger, P.~Fischer, and T.~Brox.
\newblock U-{Net}: {Convolutional} {Networks} for {Biomedical} {Image}
  {Segmentation}.
\newblock \emph{arXiv:1505.04597 [cs]}, 2015.

\bibitem[Saltikoff et~al.(2019)]{saltikoff_opera_2019}
E.~Saltikoff et~al.
\newblock {OPERA} the {Radar} {Project}.
\newblock \emph{Atmosphere}, 10\penalty0 (6):\penalty0 320, 2019.
\newblock \doi{10.3390/atmos10060320}.

\bibitem[Schneider et~al.(2022)]{esaecmwfreport2021}
R.~Schneider et~al.
\newblock {ESA-ECMWF} report on recent progress and research directions in
  machine learning for {Earth} {System} observation and prediction.
\newblock \emph{npj Clim Atmos Sci}, \penalty0 (51), 2022.
\newblock \doi{10.1038/s41612-022-00269-z}.

\bibitem[Schultz et~al.(2021)]{schultz_can_2021}
M.~G. Schultz et~al.
\newblock Can deep learning beat numerical weather prediction?
\newblock \emph{Phil. Trans. R Soc. A: Math., Phys. Eng. Sc}, 379\penalty0
  (2194):\penalty0 20200097, 2021.
\newblock \doi{10.1098/rsta.2020.0097}.

\bibitem[Sun et~al.(2014)]{sun_use_2014}
J.~Sun et~al.
\newblock Use of {NWP} for nowcasting convective precipitation: Recent progress
  and challenges.
\newblock \emph{BAMS}, 95\penalty0 (3):\penalty0 409 -- 426, 2014.
\newblock \doi{https://doi.org/10.1175/BAMS-D-11-00263.1}.

\bibitem[Sønderby et~al.(2020)]{sonderby_metnet_2020}
C.~K. Sønderby et~al.
\newblock {MetNet}: {A} {Neural} {Weather} {Model} for {Precipitation}
  {Forecasting}.
\newblock \emph{arXiv:2003.12140}, 2020.

\end{thebibliography}

%\begin{thebibliography}{00}
%\bibitem{b1} A. Author, and B. Author, ``Article presented in the conference,'' "Proc. of the %2017 conference on Big Data from Space (BiDS'17), Publications Office of the European Union, %doi: \href{http://dx.doi.org/10.2760/383579}{10.2760/383579}, 2017. 
%\end{thebibliography}

\end{document}